\documentclass[twocolumn]{aastex62}
\usepackage{amsmath}

\usepackage{float}
\usepackage{booktabs}
\graphicspath{{./}{figures/}}

\shorttitle{The Copernican Principle Rules Out A Technological Civilization in the $\mathrm{\alpha}$ Cen System}
\shortauthors{Siraj \& Loeb}

\begin{document}
\title{The Copernican Principle Rules Out BLC1 as a Technological Radio Signal from the Alpha Centauri System}

\email{amir.siraj@cfa.harvard.edu, aloeb@cfa.harvard.edu}

\author{Amir Siraj}
\affil{Department of Astronomy, Harvard University, 60 Garden Street, Cambridge, MA 02138, USA}

\author{Abraham Loeb}
\affiliation{Department of Astronomy, Harvard University, 60 Garden Street, Cambridge, MA 02138, USA}




\begin{abstract}
Without evidence for occupying a special time or location, we should not assume that we inhabit privileged circumstances in the Universe. As a result, within the context of all Earth-like planets orbiting Sun-like stars, the origin of a technological civilization on Earth should be considered a single outcome of a random process. We show that in such a Copernican framework, which is inherently optimistic about the prevalence of life in the Universe, the likelihood of the nearest star system, Alpha Centauri, hosting a radio-transmitting civilization is $\sim 10^{-8}$. This rules out, \textit{a priori}, Breakthrough Listen Candidate 1 (BLC1) as a technological radio signal from the Alpha Centauri system, as such a scenario would violate the Copernican principle by about eight orders of magnitude. We also show that the Copernican principle is consistent with the vast majority of Fast Radio Bursts being natural in origin.

\end{abstract}



\section{Introduction}
The Copernican principle asserts that we are not privileged observers of the Universe. Successes of its application include the rejection of Ptolemaic geocentrism and the adoption of the modern cosmological principle, which underpins the leading $\Lambda \mathrm{CDM}$ model \citep{1993ppc..book.....P}. By definition, there are fewer special than unspecial states in the Universe. Therefore, without evidence for occupying a special time or location, we should not assume that we inhabit privileged circumstances in the Universe. As a result, within the context of all Earth-like planets orbiting Sun-like stars, the origin of life on Earth should be considered a single outcome of a random process.

A habitable Earth around a Sun-like star is common based on Kepler satellite data analyzed by \cite{2020arXiv201014812B}, hence the dice have been rolled billions of times in the Milky Way alone. \cite{1993Natur.363..315G} applied the Copernican principle to lifetime estimation, including for the human species. \cite{2012PNAS..109..395S} conducted a Bayesian analysis on the emergence of life on Earth, however their model suffers from sensitivity to arbitrary assumptions about the boundary conditions of the prior distribution for the likelihood of the emergence of life. \cite{2019AsBio..19...28L} estimated the relative likelihood of searches for primitive and intelligent life, using a Drake-type approach. \cite{2020ApJ...896...58W} applied the Copernican principle to the search for intelligent life, but in forms that featured strict boundaries in time, thereby not reflecting a truly random process.

Here, we use a Copernican framework that operates solely on the assumption that we are not special, to derive a probability distribution for the likelihood of a radio-transmitting civilization developing around a Sun-like star. We investigate the implications of the results for Breakthrough Listen Candidate 1 (BLC1)\footnote{https://sites.psu.edu/astrowright/2020/12/20/blc1-a-candidate-signal-around-proxima/}, a radio signal detected from the apparent direction of the Alpha Centauri star system. In Section \ref{mc}, we present notation for the Copernican framework adopted here. In Section \ref{pp}, we explore the role of Poisson statistics in quantifying the Copernican principle. In Section \ref{moc}, we describe the complete Monte Carlo method for applying the Copernican principle to the prevalence of radio-transmitting civilizations. In Section \ref{ext}, we consider the implications of our model for extragalactic technosignatures. Finally, in Section \ref{d}, we explore key predictions of our model.


\section{Formalism}
\label{mc}

Consider a technosignature of life, such as radio wave transmission, on Earth-like planets orbiting Sun-like stars, with some mean emergence timescale, $\tau_{begin}^{\mu}$, and some mean extinction timescale, $\tau_{end}^{\mu}$, where $\tau_{end}^{\mu} > \tau_{begin}^{\mu}$, $\tau_{end}^{\mu} - \tau_{begin}^{\mu} = \tau_{obs}^{\mu}$, and $\tau_{obs}^{\mu} \ll \tau_{begin}^{\mu} \sim \tau_{end}^{\mu}$. A technosignature can only develop around any given star if $\tau_{begin}^{\mu} < \tau_{\star}$, where $\tau_{\star} \sim 5.5 \mathrm{\; Gyr}$ is the timescale on which Earth-like planets in the habitable zones of Sun-like stars remain unaffected by inevitable runaway greenhouse heating due to the expansion of the parent star \citep{2008MNRAS.386..155S}.

If $P_{life} = n(\{\tau_{begin} \; \vert \; \tau_{begin} < \tau_{\star}\}) \; / \; n(\{\tau_{begin}\})$ is the probability that a Sun-like star develops a technosignature within $\tau_{\star}$, and $P_{obs} = \tau_{obs}^{\mu} / \tau_{\star}$ is the probability that a Sun-like star that develops a technosignature at some time within $\tau_{\star}$ presently hosts a technosignature, then the probability $P$ that a Sun-like star with a similar age to the Sun presently hosts a technosignature is simply, $P = P_{life} \times P_{obs}$. By definition, $P$ is equivalent to the probability that any Sun-like star presently hosts a technosignature.

\section{Poisson Process}
\label{pp}

Given that the emergence and extinction of a technosignature around the Sun is both independent of the emergence and extinction of life around any other star, the average rates of emergence and extinction, $\Gamma_{begin}^{\mu} = 1 / \tau_{begin}^{\mu}$ and $\Gamma_{end}^{\mu} = 1 / \tau_{end}^{\mu}$, the emergence and extinction of technosignatures on Earth can be modeled as a Poisson process. For a given time window during which the average number of events is $\mu$, the probability of $x$ events occurring is \citep{1986ApJ...303..336G},

\begin{equation}
    P(x, \mu) = \frac{e^{-\mu} \mu^x}{x!} \; \; \; ,
\end{equation}
so the confidence level $\mathcal{P}$ that fewer than $n$ events take place is,
\begin{equation}
    \label{lowerl}
    \mathcal{P} = \sum_{x = 0}^{n-1} \; P(x, \mu) \; \; ,
\end{equation}
and the confidence level $\mathcal{P}$ that at least $n$ events take place is,
\begin{equation}
    \label{upperl}
    \mathcal{P} = 1 - \sum_{x = 0}^{n-1} \; P(x, \mu) \; \; .
\end{equation}
Solving Equation \eqref{lowerl} yields the normalized average rate, $\mu$, at percentile level $\mathcal{P}$ of the Poisson distribution for fewer than $n$ events taking place. For the case of $n = 1$,
\begin{equation}
    \label{solvelowerl}
    \mu = \ln{(\mathcal{P}^{-1})} \; \; \; .
\end{equation}
Similarly, solving Equation \eqref{upperl} yields the normalized average rate, $\mu$, at percentile level $\mathcal{P}$ of the Poisson distribution for at least $n$ events taking place. For the case of $n = 1$,
\begin{equation}
    \label{solveupperl}
    \mu = \ln{((1 - \mathcal{P})^{-1})} \; \; \; .
\end{equation} We note that, if $\mathcal{P}$ is sampled repeatedly from a uniform distribution over the interval, $[0, 1)$, Equations \eqref{solvelowerl} and \eqref{solveupperl} yield identical distributions.

Given a technosignature on Earth with observed emergence timescale, $\tau_{begin}^{\oplus}$, and observed extinction timescale, $\tau_{end}^{\oplus}$, at percentile level $\mathcal{P}$ of the respective Poisson distributions, the mean emergence timescale is,
\begin{equation}
\label{meanbegin1}
\tau_{begin}^{\mu} = (\tau_{begin}^{\oplus} / \ln{((1 - \mathcal{P})^{-1})}) \; \; \; ,
\end{equation}
and the mean extinction timescale is,
\begin{equation}
\label{meanend1}
\tau_{end}^{\mu} = \tau_{begin}^{\mu}  + (\tau_{obs}^{\oplus} / \ln{(\mathcal{P}^{-1})}) \; \; \; .
\end{equation}
By definition, for any observed technosignature on Earth, $\tau_{begin}^{\oplus} \sim 4.5 \mathrm{\; Gyr}$.

\section{Monte Carlo Model}
\label{moc}

For radio transmission in particular, the canonical technosginature, $\left( \tau_{present} - \tau_{begin}^{\oplus} \right) = 132 \mathrm{\; yr}$. This raises the question, what value should be adopted for $\tau_{obs}^{\oplus}$, if a given technosignature still exists today? We adopt the Copernican lifetime estimation method \citep{1993Natur.363..315G}, a framework with no dependence on assumed boundary conditions, in which the expected survival timescale from the present in units of the elapsed time (observed age) at some percentile level $\mathcal{P}$ is,
\begin{equation}
    \label{tauobs}
    \tau_{obs}^{\oplus} / (\tau_{present} - \tau_{begin}^{\oplus}) = (\mathcal{P}^{-1} - 1)^{-1} \; \; .
\end{equation}
We then use the following method to determine a singular stochastic instance of $P$, the probability that an Sun-like star formed $\lesssim \mathrm{5.5 \; Gyr}$ ago\footnote{The age at which the Earth will experience runaway greenhouse heating and cross the inner edge of the Sun's habitable zone \citep{2008MNRAS.386..155S}.} currently hosts radio technosignature on an Earth-like planet. We draw three independent values for percentile level $\mathcal{P}$ from a uniform distribution over the interval, $[0, 1)$, the prior distribution that adopts no artificial limit truncations. One value of $\mathcal{P}$ is used to compute a possible instance of $\tau_{begin}^{\mu}$ using Equation \eqref{meanbegin1}. Another value of $\mathcal{P}$ is used to compute a possible instance of $\tau_{obs}^{\oplus}$ using Equation \eqref{tauobs}. The final value of $\mathcal{P}$ is used to compute a possible instance of $\tau_{end}^{\mu}$ using Equation \eqref{meanend1} and the chosen value of $\tau_{obs}^{\oplus}$. Given the value of $\tau_{begin}^{\mu}$, Poisson statistics dictate that the proportion of stars with at least one emergence of a technosignature within $\tau_{\star}$ is,
\begin{equation}
    P_{life} = 1 - e^{-(\tau_{\star} / \tau_{begin}^{\mu})} \; \; ,
\end{equation}
where each instance of $\tau_{\star}$ is drawn from a triangular distribution with range $0 - 5.5 \mathrm{\; Gyr}$, an order-of-magnitude approximation of the star formation history (SFH) for stars with $\tau_{\star} \lesssim 5.5 \mathrm{\; Gyr}$ in the Solar neighborhood (Figure 19, \citealt{2021MNRAS.501..302A}) over the timescale that the Earth is located within the habitable zone of the Sun \citep{2008MNRAS.386..155S}. We additionally note that only $\sim 20 \%$ of stars were formed within the past $5.5 \mathrm{\; Gyr}$ \citep{2021MNRAS.501..302A}.

Since $P_{obs} = \tau_{obs}^{\mu} / \tau_{\star}$, the probability that a radio technosignature is presently visible from an Sun-like star, given that a radio technosignature has existed within $\tau_{\star}$,
\begin{equation}
    P = (\tau_{obs}^{\mu} / \tau_{\star}) \cdot (1 - e^{-(\tau_{\star} / \tau_{begin}^{\mu})}) \cdot n_{EL} \; \; ,
\end{equation}
where $\tau_{obs}^{\mu}$, $\tau_{\star}$, and $\tau_{begin}^{\mu}$ are stochastically determined via the aforementioned methods, and where the number of Earth-like planets per Sun-like star, $n_{EL}$, is drawn for $50\%$ of runs from the conservative habitable zone (HZ) lower case distribution ($n_{EL} = 0.36^{+0.48}_{-0.21}$) and the other $50\%$ of runs from the upper case distribution ($n_{EL} = 0.60^{+0.90}_{-0.36}$), where the quoted uncertainties represent $1-\sigma$ deviations \citep{2020arXiv201014812B}.

\begin{figure}
 \centering
\includegraphics[width=1\linewidth]{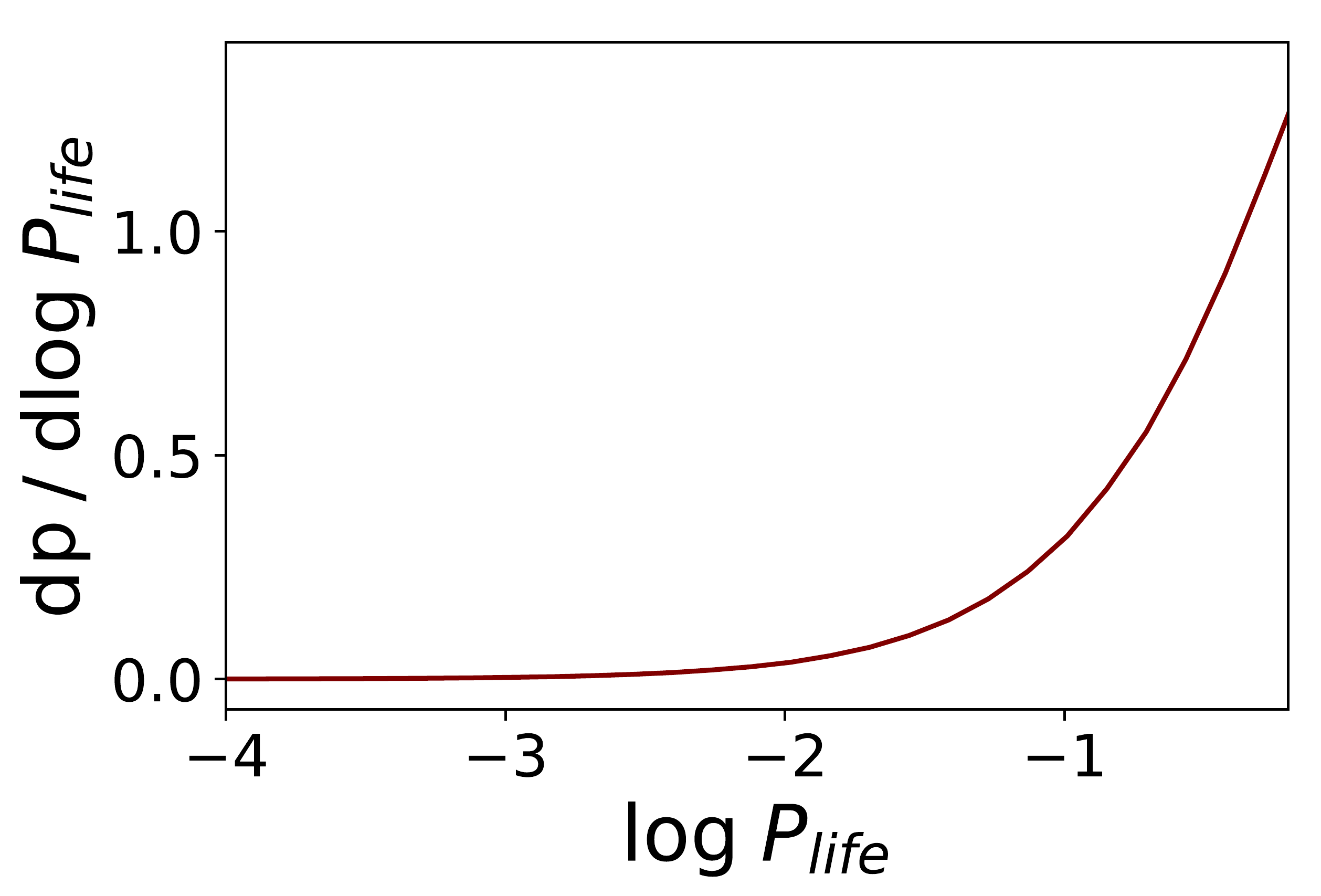}

\caption{Differential distribution of the logarithm of the probability $P_{life}$ that any given Sun-like star formed within the past $5.5 \mathrm{\; Gyr}$ has hosted a radio-transmitting civilization at some point during its lifetime ($\mathrm{dP / d \log{}} P_{life}$), derived through $10^7$ runs the Monte Carlo method described in Section \ref{moc}.}
\label{fig:plife_dist}
\end{figure}

\begin{figure}
 \centering
\includegraphics[width=1\linewidth]{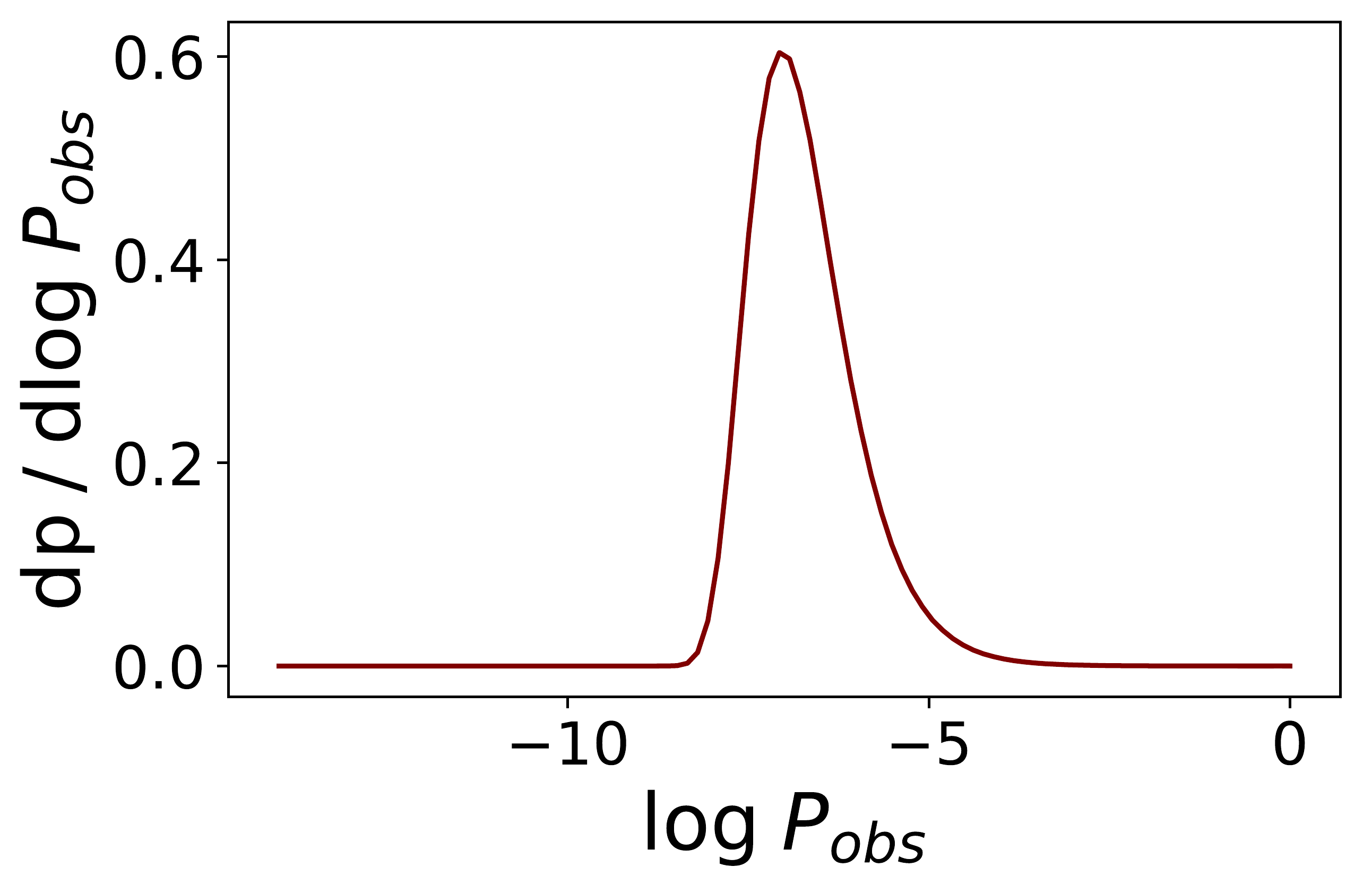}

\caption{Differential distribution of the logarithm of the probability $P_{obs}$ that any given Sun-like star formed within the past $5.5 \mathrm{\; Gyr}$ presently hosts a radio-transmitting civilization, given that it hosts one at some point during its lifetime ($\mathrm{dP / d \log{}} P_{obs}$), derived through $10^7$ runs the Monte Carlo method described in Section \ref{moc}.}
\label{fig:pobs_dist}
\end{figure}

\begin{figure}
 \centering
\includegraphics[width=1\linewidth]{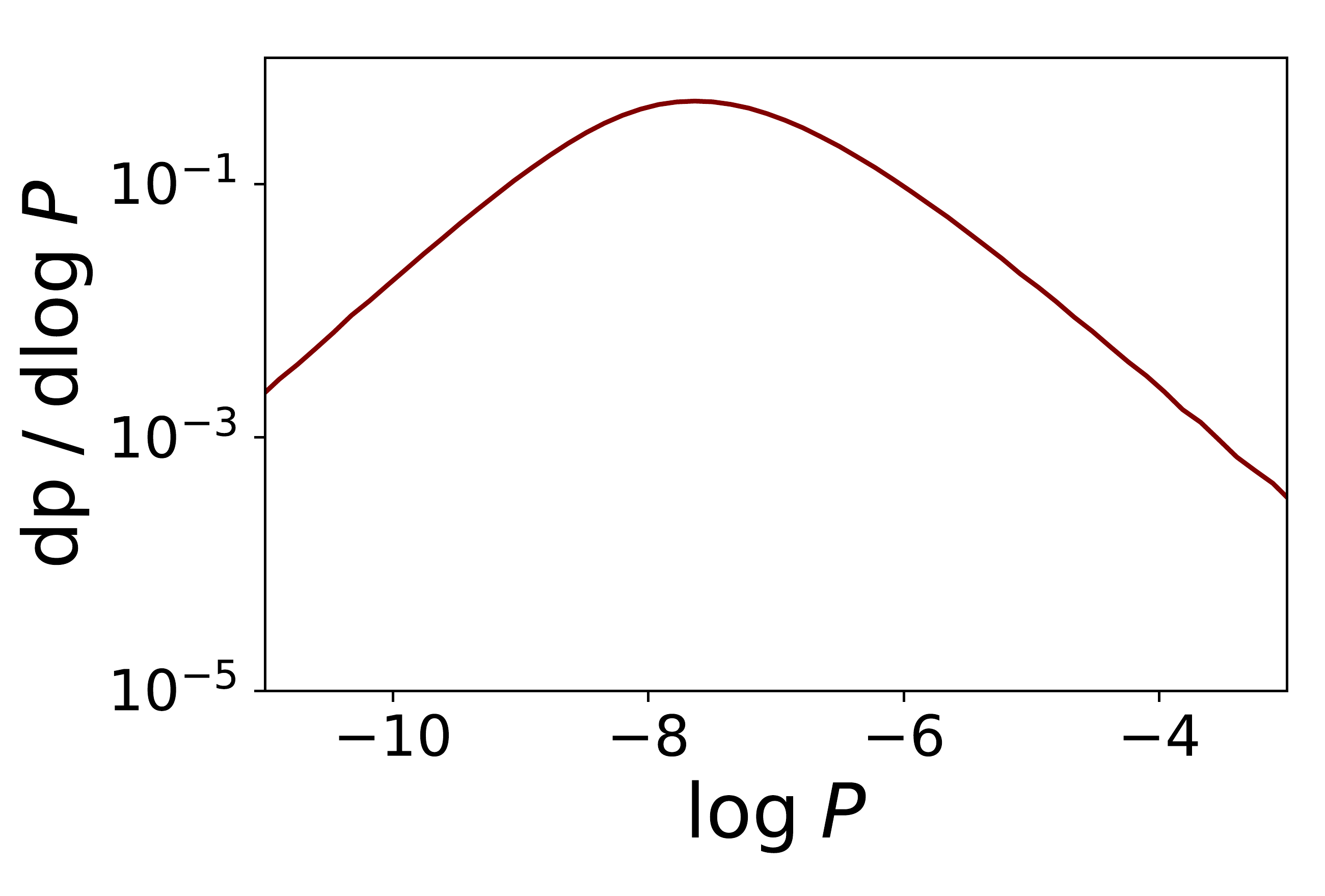}

\caption{Differential distribution of the logarithm of the probability $P = P_{life} \times P_{obs}$ that any given Sun-like star formed within the past $5.5 \mathrm{\; Gyr}$ presently hosts a radio-transmitting civilization ($\mathrm{dP / d \log{}} P$), derived through $10^7$ runs the Monte Carlo method described in Section \ref{moc}.}
\label{fig:p_dist}
\end{figure}

Figures \ref{fig:plife_dist} - \ref{fig:p_dist} display the probability distributions resulting from $10^7$ runs of the above Monte Carlo method for $P$, $P_{life}$, and $P_{obs}$. The median value of $P$ is $\sim 3 \times 10^{-8}$, and because only a fraction ($\sim 20\%$) of stars were formed within the past $5.5 \mathrm{\; Gyr}$, to order-of-magnitude the probability that any given Sun-like star presently hosts a radio transmitting civilization is $\sim 10^{-8}$. This is, as expected, of order the technological age of human civilization divided by the age of the Sun.

\newpage 
\section{Extragalactic Technosignatures}
\label{ext}

For some electromagnetic technosignature, $P \propto r^2 \nu^{-1} S_{\nu} \eta D^{-2} (\Delta \nu / \nu)$, where $P$ is power, $r$ is the distance of the source from the observer, $\nu$ is frequency, $S_{\nu}$ is the signal strength, $\eta$ is the beam diameter in units of the diffraction limit, and $D$ is the diameter of the transmitter \citep{2017ApJ...837L..23L}. Since $\eta \propto \theta \nu D$, where $\theta$ is the angular size of the beam, we can express the original relation as, $P \propto r^2 \nu S_{\nu} \theta^2 (\Delta \nu / \nu)$.

\begin{align}
\begin{split}
\label{et}
     P \sim {\mathrm{10^{25} \; erg \; s^{-1}}} & \left( \frac{r}{\mathrm{3 \; Gpc}} \right)^2 \left( \frac{\nu}{\mathrm{1 \; GHz}} \right) \left( \frac{S_{\nu}}{\mathrm{10 \; mJy}} \right)^{-1} \\ & \left( \frac{\theta}{\mathrm{10^{-5} \; rad}} \right)^{2} \left( \frac{(\Delta \nu / \nu)}{\mathrm{10^{-5}}} \right) \; \; ,
\end{split}
\end{align}
where $S_{\nu} \sim 10 \; \mathrm{mJy}$ and $\theta \sim 10^{-5} \; \mathrm{rad}$ would match the characteristic fluence of observed Fast Radio Bursts (FRBs), $(S_{\nu} \theta / \dot{\theta}) \sim 1 \; \mathrm{mJy \; s}$, where $\dot{\theta} \sim 10^{-4} \; \mathrm{rad}$ is the rotation rate of the Earth, taken to be typical of Earth-like planets.

For the fiducial parameters adopted Equation \eqref{et}, the rate of detectable signals at Earth is, $\Gamma \sim 1 \; \mathrm{yr^{-1}} \; (\mathrm{n_T / 10^{-3}})$, where $n_T$ is the number of transmitters per galaxy. Since there are $\sim 10^{10}$ Milky Way size galaxies and a few $\times \; 10^{10}$ Sun-like stars per galaxy \citep{2017ApJ...837L..23L}: according to the Copernican principle, which is consistent with $\sim 10^{-8}$ of all young Sun-like stars presently hosting radio-transmitting civilizations on Earth-like planets, even if all radio-transmitting civilizations harness the total energy of starlight reaching a habitable-zone rocky planet transmit all of that energy through a maser for a comparable duration to the time that their civilization was radio-transmitting, technosignatures would only constitute at maximum a tenth of all observed FRBs \citep{2019A&ARv..27....4P}. As a result, we find that the Copernican principle is consistent with the vast majority of FRBs being natural in origin.

\section{Discussion}
\label{d}

Here, we applied the Copernican principle to an order-of-magnitude estimate of the prevalence radio-transmitting civilizations, without boundary assumptions that have influenced some previous Copernican frameworks \citep{2012PNAS..109..395S, 2020ApJ...896...58W}, and showed that BLC1 as a candidate for a radio transmission signal from the Alpha Centauri system would be in violation of the Copernican principle by about eight orders of magnitude, since the median of the derived probability distribution for any given Sun-like star presently hosting a radio-transmitting civilization is $\sim 10^{-8}$ (see Figure \ref{fig:p_dist}). The only caveat to our conclusion is if technological life on Earth and the nearest stars is correlated, for example if the seeds for it were planted by panspermia (natural or directed) at the same time \citep{2018ApJ...868L..12G}. Here, we ignore this possibility, because \textit{Homo sapiens} appeared on Earth $\sim 0.3 \mathrm{\; Myr}$ ago \citep{2017Natur.546..289H, 2017Natur.546..293R}, before Alpha Centauri became our nearest star system.

The only plausible way to avoid violating the Copernican principle here would be to invoke some \textit{a priori} argument that falls outside of the purview of the Copernican principle, such as panspermia between the Earth and the Alpha Centauri system. We also showed that the Copernican principle is consistent with the vast majority of FRBs being natural in origin. 

We note that searching for signatures of dead civilizations\footnote{https://blogs.scientificamerican.com/observations/how-to-search-for-dead-cosmic-civilizations/}, such as discarded debris, self-replicating machines, or pollution, would be advantageous from a Copernican perspective on three counts. First, because the primary reason why the Copernican principle results in a pessimistic outlook on the prevalence of radio-transmitting civilizations is the fact that humanity has only been radio-transmitting for a very small fraction of the Earth's lifetime, implying that even if the time-integrated abundance of radio-transmitting civilizations is high, a snapshot at any given point in time would yield very few. Signatures of dead civilizations, such as physical debris, could be longer-lasting by nature. Second, a few, highly advanced, civilizations could in principle produce an amount of debris that exceeds the sum of that produced by all other civilizations, meaning that even if such civilizations are extremely rare, as they would have to be in order to remain consistent with the Copernican principle, they could still dominate the overall technosignature budget. Third, self-replicating machines\footnote{http://www.molecularassembler.com/KSRM.htm} may produce a large population of relics out of a small number of seeds.

\section*{Acknowledgements}
This work was supported in part by a grant from the Breakthrough Prize Foundation. 






\bibliography{bib}{}

\begin{thebibliography}{}
\expandafter\ifx\csname natexlab\endcsname\relax\def\natexlab#1{#1}\fi
\providecommand{\url}[1]{\href{#1}{#1}}

\bibitem[{{Alzate} {et~al.}(2021){Alzate}, {Bruzual}, \&
  {D{\'\i}az-Gonz{\'a}lez}}]{2021MNRAS.501..302A}
{Alzate}, J.~A., {Bruzual}, G., \& {D{\'\i}az-Gonz{\'a}lez}, D.~J. 2021,
  \mnras, 501, 302

\bibitem[{{Bryson} {et~al.}(2020){Bryson}, {Kunimoto}, {Kopparapu}, {Coughlin},
  {Borucki}, {Koch}, {Silva Aguirre}, {Allen}, {Barentsen}, {Batalha},
  {Berger}, {Boss}, {Buchhave}, {Burke}, {Caldwell}, {Campbell}, {Catanzarite},
  {Chandrasekharan}, {Chaplin}, {Christiansen}, {Christensen-Dalsgaard},
  {Ciardi}, {Clarke}, {Cochran}, {Dotson}, {Doyle}, {Seperuelo Duarte},
  {Dunham}, {Dupree}, {Endl}, {Fanson}, {Ford}, {Fujieh}, {Gautier}, {Geary},
  {Gilliland}, {Girouard}, {Gould}, {Haas}, {Henze}, {Holman}, {Howard},
  {Howell}, {Huber}, {Hunter}, {Jenkins}, {Kjeldsen}, {Kolodziejczak},
  {Larson}, {Latham}, {Li}, {Mathur}, {Meibom}, {Middour}, {Morris}, {Morton},
  {Mullally}, {Mullally}, {Pletcher}, {Prsa}, {Quinn}, {Quintana}, {Ragozzine},
  {Ramirez}, {Sanderfer}, {Sasselov}, {Seader}, {Shabram}, {Shporer}, {Smith},
  {Steffen}, {Still}, {Torres}, {Troeltzsch}, {Twicken}, {Kamal Uddin}, {Van
  Cleve}, {Voss}, {Weiss}, {Welsh}, {Wohler}, \&
  {Zamudio}}]{2020arXiv201014812B}
{Bryson}, S., {Kunimoto}, M., {Kopparapu}, R.~K., {et~al.} 2020, arXiv
  e-prints, arXiv:2010.14812

\bibitem[{{Gehrels}(1986)}]{1986ApJ...303..336G}
{Gehrels}, N. 1986, \apj, 303, 336

\bibitem[{{Ginsburg} {et~al.}(2018){Ginsburg}, {Lingam}, \&
  {Loeb}}]{2018ApJ...868L..12G}
{Ginsburg}, I., {Lingam}, M., \& {Loeb}, A. 2018, \apjl, 868, L12

\bibitem[{{Gott}(1993)}]{1993Natur.363..315G}
{Gott}, J.~R., I. 1993, \nat, 363, 315

\bibitem[{{Hublin} {et~al.}(2017){Hublin}, {Ben-Ncer}, {Bailey}, {Freidline},
  {Neubauer}, {Skinner}, {Bergmann}, {Le Cabec}, {Benazzi}, {Harvati}, \&
  {Gunz}}]{2017Natur.546..289H}
{Hublin}, J.-J., {Ben-Ncer}, A., {Bailey}, S.~E., {et~al.} 2017, \nat, 546, 289

\bibitem[{{Lingam} \& {Loeb}(2017)}]{2017ApJ...837L..23L}
{Lingam}, M., \& {Loeb}, A. 2017, \apjl, 837, L23

\bibitem[{{Lingam} \& {Loeb}(2019)}]{2019AsBio..19...28L}
---. 2019, Astrobiology, 19, 28

\bibitem[{{Peebles}(1993)}]{1993ppc..book.....P}
{Peebles}, P.~J.~E. 1993, {Principles of Physical Cosmology}

\bibitem[{{Petroff} {et~al.}(2019){Petroff}, {Hessels}, \&
  {Lorimer}}]{2019A&ARv..27....4P}
{Petroff}, E., {Hessels}, J.~W.~T., \& {Lorimer}, D.~R. 2019, \aapr, 27, 4

\bibitem[{{Richter} {et~al.}(2017){Richter}, {Gr{\"u}n}, {Joannes-Boyau},
  {Steele}, {Amani}, {Ru{\'e}}, {Fernandes}, {Raynal}, {Geraads}, {Ben-Ncer},
  {Hublin}, \& {McPherron}}]{2017Natur.546..293R}
{Richter}, D., {Gr{\"u}n}, R., {Joannes-Boyau}, R., {et~al.} 2017, \nat, 546,
  293

\bibitem[{{Schr{\"o}der} \& {Smith}(2008)}]{2008MNRAS.386..155S}
{Schr{\"o}der}, K.~P., \& {Smith}, R.~C. 2008, \mnras, 386, 155

\bibitem[{{Spiegel} \& {Turner}(2012)}]{2012PNAS..109..395S}
{Spiegel}, D.~S., \& {Turner}, E.~L. 2012, Proceedings of the National Academy
  of Science, 109, 395

\bibitem[{{Westby} \& {Conselice}(2020)}]{2020ApJ...896...58W}
{Westby}, T., \& {Conselice}, C.~J. 2020, \apj, 896, 58

\end{thebibliography}
\bibliographystyle{aasjournal}



\end{document}